\documentstyle[11pt,appb
]{article} 

\begin{document}
\title{TOPOLOGY OF 3-DIMENSIONAL SEIBERG-WITTEN THEORY%
\thanks{Presented by B.~B.\ at the Workshop: ``Gauge Theories of Gravitation'', Jadwisin, Poland, 4--10 September 1997.}
}
\author{Bogus\l aw BRODA$^{\rm a}$ \and Ma\l gorzata BAKALARSKA
\address{
Department of Theoretical Physics, University of \L\'od\'z\\
Pomorska 149/153, PL--90-236 \L\'od\'z, Poland\\
$^{\rm a}$\tt bobroda@krysia.uni.lodz.pl}
}
\maketitle
\begin{abstract}
A dual description of 3-dimensional topological Seiberg-Witten theory in terms of the Alexander invariant on manifolds obtained via surgery on a knot is proposed. The description directly follows from a low-energy analysis of the corresponding SUSY theory, in full analogy to the 4-dimensional case.
\end{abstract}

\section{Introduction}
In 1988 Witten proposed a description of Donaldson's theory (``topological'' invariants of four-dimensional manifolds) in terms of an appropriately twisted ${\cal N}=2$ SUSY SU(2) pure gauge theory \cite{w1}. But not so many mathematical consequences have followed from this approach until 1994.
Since 1994 we have witnessed a new (second) ``revolution'' in this fascinating area at the border of quantum field theory and topology known under the name of (topological) Seiberg-Witten (SW) theory. It has appeared that a newly-discovered dual description of the ${\cal N}=2$ theory in low-energy limit \cite{sw} provides us a new, alternative (but essentially equivalent and simpler) formulation of invariants of four-dimensional manifolds \cite{w2}. It would be greatly interesting and natural to investigate a presumably analogous situation in three dimensions. And this is the aim of our present work.

In a previous paper \cite{b}, a physical scenario has been proposed to reach our present goal using a geometric-topological construction with scalar curvature distribution ``compatible'' with surgery. Though the scalar curvature distribution used ``agrees'' with the surgery procedure, nevertheless it is unclear why such a distribution should be privileged. Therefore, in the present paper, we propose a simpler and more natural mechanism without any reference to curvature. In fact, we directly apply known results concerning low-energy limit of 3d ${\cal N}=4$ SUSY U(1) gauge theory with matter field \cite{isw}.

\section{Physical picture}
In three dimensions (3d), we have the two important topological quantum field theories (of ``cohomological'' type): topological SU(2) gauge theory of flat connection and 3d version of the (topological) SW theory. The former is a 3d twisted ${\cal N}=4$ SUSY SU(2) pure gauge theory or a 3d version of the Donaldson-Witten (DW) theory, and ``by definition'' it describes the Casson invariant, which appropriately counts the number of flat SU(2) connections \cite{bt}. The latter is a 3d twisted version of ${\cal N}=4$ SUSY U(1) gauge theory with a matter hypermultiplet \cite{sw}, \cite{w2}. Also this theory should describe an interesting non-trivial topological invariant of 3d manifolds pertaining to SW invariant which was conjectured to be equivalent to the Casson invariant (and thus to the former theory) or to topological torsion. The first conjecture is physically strongly motivated by the fact that the both theories can be derived from 4d ${\cal N}=2$ SUSY SU(2) pure gauge theory corresponding via twist to DW theory. The second one is a mathematical result \cite{mthl}. This equivalence would be a 3d counterpart of the equivalence of 4d DW and SW theories. The latter being a ``low-energy version'' of the former. In this paper, using physical arguments, we will show that topological contents of 3d SW theory is basically equivalent to an abelian version of the Casson invariant (of infinite covering space $\widetilde{{\cal M}^3}$). In turn, the abelian Casson invariant will appear to be equivalent to the Alexander ``polynomial'' of the manifold ${\cal M}^3$ itself (see, also \cite{fn}). In fact, as follows from mathematical literature the Alexander invariant is related to the (non-abelian) Casson invariant as well as to topological torsion \cite{t}.
The 3d manifold we are interested in, ${\cal M}^3$, is obtained from the 3d sphere ${\cal S}^3$ via 0-surgery on a knot.

To determine the dual description of SW topological invariants of the 3d manifold ${\cal M}^3$ we will make use of known results pertaining to the corresponding physical (untwisted) theory. The corresponding physical theory is 3d ${\cal N}=4$ SUSY U(1) gauge theory with a single (monopole) matter hypermultiplet. Its low-energy limit has been determined by Seiberg and Witten in \cite{isw}. The theory has only the Coulomb branch (there is no Higgs branch for one monopole field). According to \cite{isw}, the moduli space of the theory is everywhere smooth, and the monopole is nowhere massless. Thus, in low-energy limit, we deal with ${\cal N}=4$ SUSY U(1) pure gauge theory, i.e.\ 3d (super)electrodynamics. In topological sector (upon twisting), we obtain a theory of flat U(1) connection. At first sight, the theory could only trivially detect the first cohomology of ${\cal M}^3$ (not a very exciting quantity). But happily, this is not the case as the dual of the photon assumes values on a circle ${\cal S}^1$ rather than on ${\cal R}^1$ \cite{isw}, and in the case of non-simply-connected manifolds the topological theory begins to live on the infinite covering space $\widetilde{{\cal M}^3}$ of the original manifold ${\cal M}^3$, where the dual photon field would be a multivalued function. Therefore, actually we measure the first (co)homology of $\widetilde{{\cal M}^3}$ (the Alexander invariant), rather than the trivial one of ${\cal M}^3$. And more precisely, we count the number of U(1) connections on $\widetilde{{\cal M}^3}$.

That way, in topological sector, the theory has switched from measuring the first cohomology of ${\cal M}^3$ to measuring the first cohomology of its infinite cover, $\widetilde{{\cal M}^3}$. In other words, upon twisting, we have an abelian ``cohomological'' theory on $\widetilde{{\cal M}^3}$.
The partition function $Z$ of our theory should ``ordinarily'', i.e. without sign (due to abelianity), count the number of (inequivalent)
U(1)-representations of the fundamental group $\pi_1$ of $\widetilde{{\cal M}^3}$
$$
Z({\cal M}^3) \propto I(\widetilde{{\cal M}^3}) = \# \hbox{Rep} \left( \pi_1\left(\widetilde{{\cal M}^3}\right) \longrightarrow {\rm U(1)}\right).
\eqno{(1)}
$$
Eq.~(1) defines an abelian analogue of the (standard) Casson invariant. Our temporary conclusion says that the partition function $Z$ of 3d SW theory describes the abelian Casson invariant $I$ of $\widetilde{{\cal M}^3}$. Formula (1) is a counterpart of an analogous formula for topological field theory of flat SU(2) connection.

\section{Meaning of the partition function}
Now, we should translate the invariant of $\widetilde{{\cal M}^3}$ to a, possibly already known, topological invariant of ${\cal M}^3$. To this end, we will ``explicitly calculate'' the number of $\hbox{Rep} \left( \pi_1\left(\widetilde{{\cal M}^3}\right) \longrightarrow {\rm U(1)}\right)$ \cite{b}. First of all, since U(1) is an abelian group a non-abelian part of $\pi_1$ drops out, and effectively we deal with an equivalent but simpler expression,
$$
I(\widetilde{{\cal M}^3})=\# \hbox{Rep} \left( H_1\left(\widetilde{{\cal M}^3}\right) \longrightarrow {\rm U(1)}\right),
\eqno{(2)}
$$
where $H_1$ is an integer-valued first homology group ($=\pi_1/\left[\pi_1,\pi_1\right]$). We claim that $I$ is the Alexander ``polynomial'' understood as a determinant
$$
I(\widetilde{{\cal M}^3})=\det A_{kl},
\eqno{(3)}
$$
where the matrix $A$ ``describes'' the homology of $\widetilde{{\cal M}^3}$
$$
\sum_l A_{kl} \alpha_l=0,
\qquad
k,l=1,\dots,N,
\eqno{(4)}
$$
for $\alpha_l$---some homology basis (see, \cite{r}). Now, we should pass to a U(1) representation of (4). If
$$
\alpha_l \longrightarrow e^{2\pi i \omega_l},
$$
then for LHS of (4) we have
$$
\sum_l A_{kl} \alpha_l \longrightarrow \prod_l \exp\left(2\pi i A_{kl} \omega_l\right)
= \exp\left(2\pi i \sum_l A_{kl} \omega_l\right),
$$
where $0\leq\omega_1,\dots,\omega_N<1$, for uniqueness. RHS of Eq.~(4) is now
$$
0\longrightarrow 1 = \exp(2\pi i m_k),
\qquad
m_k \in Z.
$$
A new, U(1) version of Eq.~(4) is then
$$
\sum_l A_{kl} \omega_l = m_k.
\eqno{(5)}
$$
By virtue of (2), $I$ is the number of solutions of Eq.~(5), i.e.\ the number of different $m_k$. $m_k$ are integer-valued points in a parallelepiped spanned by the ``base vectors'' $A_{k(l)}$. One can observe that the number of $m$'s is equal to the number of points inside the parallelepiped with integer coordinates and thus to the volume of that parallelepiped, 
which in turn, is equal to the determinant of $A_{kl}$.

\section{Conclusions}
In this paper, we have proposed a physical scenario describing 3d topological SW theory in terms of the ``Alexander polynomial'' (3) of the 3d manifold ${\cal M}^3$ obtained via 0-surgery on a knot. This dual, ``low-energy'' description of 3d topology directly follows from the corresponding facts concerning physical 3d ${\cal N}=4$ SUSY U(1) gauge theory with matter. 

\section{Acknowledgments}
B.~B. is greatly indebted to the organizers of the Workshop for their kind invitation to Jadwisin. B.~B. is also grateful to Prof. C.~B\"ar for an interesting discussion.
The paper has been supported by the KBN grant 2P03B09410.

\eject


\begin{thebibliography}{xxx}

\bibitem{w1}
E. Witten, Topological quantum field theory, {\it Commun.Math.Phys.}{\bf 117}, 353(1988).

\bibitem{sw}
N. Seiberg, E. Witten, Electric-magnetic duality, monopole condensation, and confinement in N=2 supersymmetric Yang-Mills theory, {\it Nucl.Phys.B}{\bf 426}, 19(1994).

\bibitem{w2}
E. Witten, Monopoles and four-manifolds, {\it Math.Res.Lett.}{\bf 1}, 769(1994).

\bibitem{b}
B. Broda, Topological contents of 3D Seiberg-Witten theory, {\it New Developments in Quantum Field Theory} (NATO ASI Series, Zakopane, June 1997, ed. P.H. Damgaard), Plenum Press, pp. 261--268.

\bibitem{isw}
N. Seiberg, IR Dynamics on Branes and Space-Time Geometry, {\it Phys. Lett.B}{\bf 384}, 81(1996).
\\
N. Seiberg, E. Witten, Gauge Dynamics And Compactification To Three Dimensions, {\it E-print} hep-th/9607163.

\bibitem{bt}
M. Blau, G. Thompson, N=2 topological gauge theory, the Euler characteristic of moduli spaces, and the Casson invariant, {\it Commun.Math.Phys.}{\bf 152}, 41(1993).

\bibitem{mthl}
G. Meng, C. Taubes, \underbar{SW}=Milnor torsion, {\it Math.Res.Lett.}{\bf 3}, 661(1996).
\\
M. Hutchings, Y. Lee, Circle-valued Morse theory, Reidemeister torsion, and Seiberg-Witten invariants of 3-manifolds, {\it E-print} math.DG/9612004;

\bibitem{fn}
C. Frohman, A. Nicas, The Alexander polynomial via topological quantum field theory, {\it Differential Geometry, Global Analysis, and Topology} (Canadian Math.Soc.Conf.Proc., Vol.12, Amer.Math.Soc.), Providence 1992, pp. 27--40.

\bibitem{t}
V. Turaev, Reidemeister torsion in knot theory, {\it Russian Math.Surveys} {\bf 41}, 119(1986).

\bibitem{r}
D. Rolfsen, {\it Knots and Links}, Publish or Perish, Inc., Wilmington 1976.

\end{thebibliography}
\end{document}